\documentstyle[11pt,iau185,twoside,epsf]{article}

\begin{document}

\title{Constraints on $^{12}{\rm C}(\alpha ,\gamma)^{16}{\rm O}$ 
       from White Dwarf Seismology}

\author{T. S. Metcalfe\altaffilmark{1}}
\affil{Theoretical Astrophysics Center, Institute of Physics and Astronomy,
Aarhus University, 8000 Aarhus C, Denmark}
\altaffiltext{1}{Formerly at the Department of Astronomy,
University of Texas-Austin; e-mail: travis@ifa.au.dk}

\author{M. Salaris}
\affil{Astrophysics Research Institute, Liverpool John Moores University, 
Twelve Quays House, Egerton Wharf, Birkenhead CH41 1LD, UK}

\author{D. E. Winget}
\affil{Department of Astronomy (C1400), University of Texas, Austin TX
78712, USA}

\begin{abstract}
For many years, astronomers have promised that the study of pulsating
white dwarfs would ultimately lead to useful information about the physics
of matter under extreme conditions of temperature and pressure. We can now
make good on that promise. Using observational data from the Whole Earth
Telescope and a new analysis method employing a genetic algorithm, we
empirically determine the central oxygen abundance of the helium-atmosphere 
variable white dwarf \object{GD\,358}. We use this value, combined with
detailed evolutionary calculations of the internal chemical profiles to
place constraints on the $^{12}{\rm C}(\alpha,\gamma)^{16}{\rm O}$ nuclear
reaction cross-section.
\end{abstract}


\section{Introduction}

During helium burning in the core of a red giant star, only two nuclear
reactions are effectively competing for the available helium nuclei:  [1]
the triple-$\alpha$ process, which fuses three helium nuclei into carbon,
and [2] the $^{12}{\rm C}(\alpha,\gamma)^{16}{\rm O}$ reaction, which
combines the carbon with another helium nucleus to produce oxygen. The
triple-$\alpha$ reaction has been measured very precisely in the
laboratory, but the same is not true of the 
$^{12}{\rm C}(\alpha,\gamma)^{16}{\rm O}$ reaction. Our current knowledge
of its rate at stellar energies is the result of a very uncertain
extrapolation from high energy measurements. This translates into
similarly large uncertainties in our understanding of every astrophysical
process that depends on this reaction, including supernovae explosions and
galactic chemical evolution.

The relative success of the two reactions inside a red giant determines
the final mixture of carbon and oxygen in the resulting white dwarf star.  
Since the triple-$\alpha$ process is already well constrained, a
measurement of the internal composition of a white dwarf can improve our
understanding of the $^{12}{\rm C}(\alpha,\gamma)^{16}{\rm O}$ reaction.
Although we cannot see directly inside a white dwarf star, a fortunate
circumstance allows us to learn about the interior. After shedding its red
giant cocoon to form a planetary nebula, the white dwarf begins to cool.
As it fades, it will pass through one of several narrow ranges in
temperature that may induce subtle periodic vibrations, sending seismic
waves deep through the interior and bringing information to the surface in
the form of brightness variations.  Careful observations of these periodic
changes reveal patterns that can be reproduced with a fair degree of
accuracy using relatively simple computer models of white dwarf stars. By
adjusting the characteristics of the model to provide the closest possible
match to the observations, we can infer the internal composition and
structure of the actual white dwarf.

\section{Method}

In the past decade, the observational requirements of white dwarf
seismology have been satisfied by the development of the Whole Earth
Telescope---a group of astronomers distributed around the globe who
cooperate to observe these stars continuously for up to two weeks at a
time (Nather et al.\,1990). This instrument is now mature, and has
provided a wealth of seismological data on the different varieties of
pulsating white dwarf stars.

In an effort to bring the analysis of these data to the level of
sophistication demanded by the observations, we have recently developed a
new model-fitting method based on a genetic algorithm (Metcalfe 2001). The
underlying ideas for genetic algorithms were inspired by Charles Darwin's
notion of biological evolution through natural selection. The basic idea
is to solve a problem by {\it evolving} the best solution from an initial
set of random guesses.  In practice, this approach allows us to find the
global solution in a parameter-space with $10^{10}$ grid points by
performing only a few\,$\times\,10^6$ model evaluations. To complete the
calculations on a reasonable timescale, we designed and built a
specialized metacomputer to run the models in parallel (Metcalfe \& Nather
2000).

The initial application of this new method to the well-observed pulsating
white dwarf \object{GD\,358} demonstrated that the models are very
sensitive to the internal composition and structure (Metcalfe, Nather, \&
Winget 2000).  In a recent follow-up study, an extension to the method
finally yielded a preliminary constraint on the 
$^{12}{\rm C}(\alpha,\gamma)^{16}{\rm O}$ reaction (Metcalfe, Winget, \&
Charbonneau 2001). More precise constraints require additional detailed
simulations of the internal chemical profiles of white dwarfs (cf. Salaris
et al.\,1997), and a thorough investigation of the systematic
uncertainties.

\section{Results}

Metcalfe, Winget, \& Charbonneau (2001) derived a central oxygen abundance
for \object{GD\,358} of $X_{\rm O} = 84 \pm 3$ percent, with the
transition from constant oxygen beginning at $q = 0.49\pm0.01~m/M_*$. To
evaluate this result, we calculated new internal oxygen profiles for a
$0.65~M_{\sun}$ white dwarf model using the same method and code described
in Salaris et al.\,(1997), but updated to use the nuclear reaction rates
from the NACRE collaboration (Angulo et al.\,1999) rather than from
Caughlin et al.\,(1985).

\begin{figure}
\begin{center}
\mbox{\epsfxsize=0.7\textwidth\epsfysize=0.7\textwidth\epsfbox{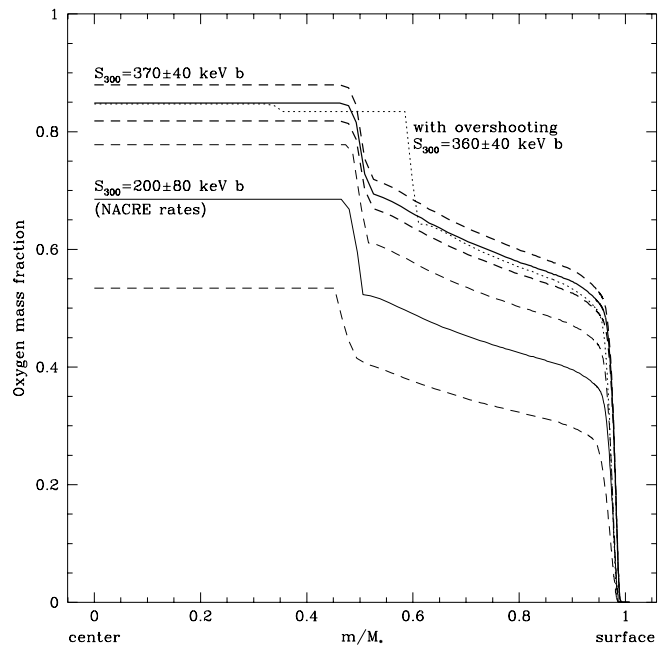}}
\caption{The internal oxygen profiles for a $0.65~M_{\sun}$ white dwarf
model using the NACRE rates (solid line), and the $\pm 1\sigma$ limits for
the $^{12}{\rm C}(\alpha,\gamma)^{16}{\rm O}$ rate (dashed lines). Also
shown are profiles using the rate that matches the central oxygen mass
fraction derived for \object{GD\,358} (dark solid line) with the $\pm
1\sigma$ limits (dark dashed lines), and when overshooting is included
(dotted line).\label{fig1}}
\end{center}
\end{figure}

The total rate of the $^{12}{\rm C}(\alpha,\gamma)^{16}{\rm O}$ reaction
at stellar energies in the NACRE compilation corresponds to an
astrophysical S factor at 300 keV of $S_{300} = 200 \pm 80$ keV-b,
yielding a central oxygen abundance between 0.53 and 0.78. To generate
profiles with a higher central oxygen abundance, we simply scaled the
value of $S_{300}$ in the simulations. Matching the central oxygen mass
fraction inferred for \object{GD\,358} required a value of $S_{300} = 370
\pm 40$ keV-b, or $360 \pm 40$ keV-b when convective overshooting was
included (see Figure \ref{fig1}).

The oxygen profiles from the simulations performed without overshooting
consistently show the transition from constant oxygen beginning near a
fractional mass of $q\sim0.5$, regardless of the assumed value for the
$^{12}{\rm C}(\alpha,\gamma)^{16}{\rm O}$ rate. This is in good agreement
with the value for $q$ found by Metcalfe, Winget, \& Charbonneau (2001)
even though they did not use evolutionary profiles. With convective
overshooting included ($\alpha=0.20~H_p$), this transition moves out to
$q\sim0.6$ and the shape of the profile is otherwise similar.

\section{Future Work}

We are now working to quantify the systematic errors of this method by
repeating both the model-fitting procedure and the chemical profile
simulations with different assumptions. Early results suggest that these
uncertainties are not much larger than the internal errors. We also have
plans to apply this method to additional DBV stars as new data becomes
available.

\acknowledgements We would like to thank Ed Nather and Paul Charbonneau
for helpful discussions. This work was supported by grant NAG5-9321 from
the Applied Information Systems Research Program of NASA, and in part by
grant AST-9876730 from the NSF.

\section*{Discussion}

{\it J. Christensen-Dalsgaard~:} How sensitive are your results to
uncertainties in the physics, e.g., the equation of state?\\[0.2cm]
{\it T. Metcalfe~:} We are in the process of quantifying the
model-dependent uncertainties due to the physical assumptions built into
both the white dwarf pulsation code and the models of the internal
chemical profiles. At this time I can tell you that the systematic errors
are comparable to the internal errors.\\[0.4cm]
{\it G. Handler~:} Isn't there some chance that the ``best'' solution
gives unrealistic results? I was somewhat surprised by the high He-layer
mass you found.\\[0.2cm]
{\it T. Metcalfe~:} We were also initially surprised by the He-layer mass,
but it provides a much better fit to the data. We have applied the fitting
method to simulated data, and the probability that we have not found the
optimal solution is extremely small.\\[0.4cm]
{\it H. Shibahashi~:} Could you briefly tell us the advantages and the
disadvantages of the genetic algorithm?\\[0.2cm]
{\it T. Metcalfe~:} This optimization method is global, objective, and
very efficient compared to a full grid search. However, it is still fairly
computationally intensive. If you want the global solution, you have to
pay for it---the genetic algorithm simply gives you a substantial
discount.

\end{document}